\documentclass[a4paper,aip,apl,amsmath,amssymb,reprint,superscriptaddress]{revtex4-1}

\usepackage{color}

\usepackage{graphicx}
\usepackage{mathtools}
\usepackage{verbatim}
\usepackage{amsmath}
\usepackage{caption}
\usepackage{subcaption}
\usepackage{appendix}
\usepackage{esint}
\usepackage{soul}
\usepackage{dcolumn}
\usepackage{bm}
\usepackage{bbold}
\usepackage{color}
\usepackage{amsfonts}
\usepackage{amssymb}
\usepackage{mathrsfs}
\usepackage{tabularx}
\usepackage{braket}

\begin{document}

\title{Modified Axion Electrodynamics as Impressed Electromagnetic Sources Through Oscillating Background Polarization and Magnetization}

\author{Michael E. Tobar}
\email[]{michael.tobar@uwa.edu.au}
\affiliation{ARC Centre of Excellence For Engineered Quantum Systems, Department of Physics, School of Physics, Mathematics and Computing, University of Western Australia, 35 Stirling Highway, Crawley WA 6009, Australia.}
\author{Ben T. McAllister}
\affiliation{ARC Centre of Excellence For Engineered Quantum Systems, Department of Physics, School of Physics, Mathematics and Computing, University of Western Australia, 35 Stirling Highway, Crawley WA 6009, Australia.}
\author{Maxim Goryachev}
\affiliation{ARC Centre of Excellence For Engineered Quantum Systems, Department of Physics, School of Physics, Mathematics and Computing, University of Western Australia, 35 Stirling Highway, Crawley WA 6009, Australia.}

\date{\today}

\begin{abstract}
A reformulation of axion modified electrodynamics is presented where the equations maintain a similar form to the unmodified Maxwell's equations, with all modifications redefined within the constitutive relations between the $\vec{D}$, $\vec{H}$, $\vec{B}$ and $\vec{E}$ fields. This allows the interpretation of the axion induced background bound charge, polarization current and bound current along with the axion induced polarization and magnetization with the former satisfying the charge-current continuity equation. This representation is of similar form to odd-parity Lorentz invariance violating background fields in the photon sector of the Standard Model Extension. We show that when a DC $\vec{B}$-field is applied an oscillating background polarization is induced at a frequency equivalent to the axion mass. In contrast, when a large DC $\vec{E}$-field is applied, an oscillating background magnetization is induced at a frequency equivalent to the axion mass. It is evident that these terms are equivalent to impressed source terms, analogous to the way that voltage and current sources are impressed into Maxwell's equations in circuit and antenna theory. The impressed source terms represent the conversion of external energy into electromagnetic energy due to the inverse Primakoff effect converting energy from axions into oscillating electromagnetic fields. It is shown that the impressed electrical DC current that drives a DC magnetic field of an electromagnet, induces an impressed effective magnetic current (or voltage source) parallel to the DC electrical current oscillating at the Compton frequency of the axion. The effective magnetic current drives a voltage source through an electric vector potential and also defines the boundary condition of the oscillating axion induced polarization (or impressed axion induced electric field) inside and outside the electromagnet. This impressed electric field, like in any voltage source, represents an extra force per unit charge supplied to the system, which also adds to the Lorentz force.

\end{abstract}


\maketitle

\section*{Introduction}
Axions are neutral spin-zero bosons, which were originally postulated to solve the strong charge-parity problem in QCD~\cite{PQ1977,Wilczek1978,Weinberg1978,wisps}. Following this came the realization that axions could have been abundantly produced during the QCD phase transition in the early universe, and that they may be formulated as cold dark matter~\cite{Preskill1983,Sikivie1983,Sikivie1983b,Dine1983}. If this is true then the QCD axion should be abundant in the laboratory frame on earth, and thus in principle detectable on earth. Models predict that axions couple to photons, with the strength of this axion-photon interaction, and the mass of the axion given by
\begin{equation}
	g_{a\gamma\gamma}=\frac{g_\gamma\alpha}{f_{a}\pi},~~~~~~~
	m_a=\frac{z^{1/2}}{1+z}~\frac{f_{\pi}m_{\pi}}{f_a}.
	\label{eq:basic}
\end{equation}
Here $z$ is the ratio of up and down quark masses, $\frac{m_u}{m_d}~\approx$ 0.56, $f_\pi$ is the pion decay constant $\approx$ 93 MeV, $m_\pi$ is the neutral pion mass $\approx$ 135 MeV, $g_\gamma$ is an axion-model dependent parameter of order 1, and $\alpha$ is the fine structure constant~\cite{K79,Kim2010,Zhitnitsky:1980tq,DFS81,SVZ80,Dine1983}. 

The coupling of photons to axions modifies Maxwell's equations, with extra terms appearing related to the axion-photon interaction. In this work we reorganise the modified equations and define the axion induced background magnetization and polarization fields from the extra terms, which oscillate at a frequency dependent on the axion mass. We show that these fields may be considered as impressed magnetic and electric fields respectively, which contribute to the Lorentz force and are sourced from impressed charges and currents depending on the applied magnetic and electric fields and the axion amplitude and coupling. Following this we derive the modifications to the electromagnetic boundary conditions, and then apply the two potential technique and the quasi-static technique to determine the dominant fields at low-mass values. The results are then applied to an infinite solenoid, a structure commonly considered in axion detection experiments. 

A key result is that the inverse Primakoff effect creates an effective impressed magnetic current oscillating at the Compton frequency of the axion when a DC electrical current is impressed as a source term of the modified equations (necessary to create a large DC magnetic field). It is shown that the impressed effective magnetic current represents the conversion of axion energy to electromagnetic energy via a voltage source oscillating at the axions Compton frequency.

\section{Axion Modified Electrodynamics}

Axion modified electrodynamics (or axion modified Maxwell's equations) are derived from the Lagrangian for axions coupled to photons~\cite{Wilczek:1987aa}:
\begin{equation}
\mathcal{L} = \frac{1}{2}(\partial_\mu a)^2-\frac{1}{2}m{_a}^2a^2-\frac{1}{4}F_{\mu\nu}F^{\mu\nu}+\frac{1}{4}g_{a\gamma\gamma}aF_{\mu\nu}\tilde{F}^{\mu\nu}.
\end{equation}
From here, it can be shown that we arrive at the modified Maxwell's equations (in SI units, with source terms and magnetic and dielectric media)~\cite{AxionMaxwells}, given by
\begin{equation}
\vec{\nabla}\cdot\vec{D} =\rho_f+ g_{a\gamma\gamma} \sqrt{\frac{\epsilon_0}{\mu_0}} \vec{B}\cdot\vec{\nabla} a,\label{eq:M1}
\end{equation}
\begin{equation}
\vec{\nabla} \times\vec{H}= \vec{J_f} +\frac{\partial \vec{D} }{\partial t}-g_{a\gamma\gamma}\sqrt{\frac{\epsilon_0}{\mu_0}}\left(\vec{B}\frac{\partial a}{\partial t}+\vec{\nabla} a\times\vec{E}\right),\label{eq:M2}
\end{equation}
\begin{equation}
\vec{\nabla} \cdot \vec{B}= 0,\label{eq:M3}
\end{equation}
\begin{equation}
\vec{\nabla} \times \vec{E} = -\frac{\partial \vec{B}}{\partial t}\label{eq:M4},
\end{equation}
where the constitutive relationships are, $\vec{D}=\epsilon_0\vec{E}+\vec{P}$ and  $\vec{H}=\vec{B}/\mu_0-\vec{M}$. Here $\vec{D}$ is the electric flux density (or $\vec{D}$-field), $\vec{E}$ is electric field intensity (or irrotational $\vec{E}$-field), $\vec{P}$ is the polarization (or $\vec{P}$-field), $\rho_f$ is the free charge density, $\epsilon_0$ the vacuum permittivity, $\vec{H}$ is magnetic field intensity (or $\vec{H}$-field), $\vec{B}$ is magnetic flux density (or solenoidal $\vec{B}$-field), $\vec{M}$ is the magnetization (or $\vec{M}$-field), $\mu_0$ is the vacuum permeability, $\vec{J}_f$ is the free current density and $a$ is the axion scalar field, which is generally of the form $a(t,\vec{r})=a_0\cos(\vec{k}_a\cdot\vec{r}-\omega_at)$, where $\omega_a$ is the Compton angular frequency (equivalent to the axion mass in natural units) and $\vec{k}_a$ is the the effective wave vector related to the axion Compton wavelength. However, for most experiments it is sufficient to ignore the axion spatial dependence and assume $a(t)=a_0\cos(-\omega_at)$, or in complex form $a=a_0e^{-j\omega_at}$. Note that all terms containing $g_{a\gamma\gamma}$ are sometimes presented with the opposite sign, but this does not have an impact on this work as both representation are correct. 

Typically, when the spatial dependence is ignored it is common to set $\vec{\nabla} a=0$ in equations (\ref{eq:M1}) and (\ref{eq:M2}) leaving equation (\ref{eq:M2}) with the only axion modification, given by;
\begin{equation}
\vec{\nabla} \times\vec{H}= \vec{J_f} +\frac{\partial \vec{D} }{\partial t}+\vec{J}_{a}, \  \vec{J}_{a}=-g_{a\gamma\gamma}\sqrt{\frac{\epsilon_0}{\mu_0}}\vec{B}\frac{\partial a}{\partial t}\label{eq:M2b}
\end{equation}
Here, $\vec{J}_{a}$ is considered as an axion-sourced effective oscillating current, which oscillates at a frequency dictated by the inverse Primakoff effect. In the past, this effective current has been considered "non-interacting" \cite{ABRACADABRA,Ouellet2018}, i.e. it is assumed that only the magnetic field generated by $\vec{J}_{a}$ is detectable and the current itself creates no direct electric effect, which will act on electrons. In this work, we show that this axion effective current does indeed create a direct effect on electrons and can thus be considered as "interacting". We show that axion effective current can be considered as an oscillating displacement current (or oscillating impressed electric field), sourced by an effective axion induced AC voltage. Past literature, which implements modified axion electrodynamics to calculate the conversion of putative low-mass axions into electromagnetic energy under a magnetic field has overlooked this subtlety, due to undertaking the approximation $\vec{\nabla} a=0$ at the commencement of calculations \cite{ABRACADABRA}. Thus, taking this approximation to early hides the relationship between the modified Gauss' and Ampere's law, and we assert that this approximation should only be made at the end of the required calculations. The loss of this part of the low-mass solution occurs because the left side of these equations are represented as a divergence and curl respectively, and the right hand side can only be equated properly to the left if we represent the right hand side as a divergence and curl as well. In the following sub-section we undertake this approach and generalise the calculation of the axion effective current.

\subsection{Alternative Representation of Modified Axion Electrodynamics}

By substituting the following vector identities, $\vec{B}\cdot\vec{\nabla} a=\vec{\nabla}\cdot a\vec{B}-a(\vec{\nabla}\cdot \vec{B})$ and $\vec{\nabla}a\times\vec{E} =(\vec{\nabla} \times a\vec{E})-a(\vec{\nabla} \times \vec{E})$ along with (\ref{eq:M3}) and (\ref{eq:M4}), into equations (\ref{eq:M1}) and (\ref{eq:M2}), the modified Gauss' and Ampere's Laws become
\begin{equation}
\vec{\nabla}\cdot\vec{D} =\rho_f+ g_{a\gamma\gamma} \sqrt{\frac{\epsilon_0}{\mu_0}}\vec{\nabla}\cdot (a\vec{B}),\label{eq:M5}
\end{equation}
\begin{equation}
\vec{\nabla} \times\vec{H}= \vec{J_f} +\frac{\partial \vec{D} }{\partial t}-g_{a\gamma\gamma}\sqrt{\frac{\epsilon_0}{\mu_0}}\left(\frac{\partial (a\vec{B})}{\partial t}+\vec{\nabla}\times (a\vec{E})\right),\label{eq:M6}
\end{equation}
which is an alternative way of presenting modified axion electrodynamics. This representation presents the photon-axion interaction term as the product of the axion scalar amplitude, $a(t,\vec{r})$, multiplied by either the applied $\vec{E}$-field or $\vec{B}$-field. This is similar to the form of the equations in~\cite{VISINELLI}, but without the magnetic monopole duality.  For the more commonly used equations, Eqns.~(\ref{eq:M1})-(\ref{eq:M4}), the last term in Eqn.~(\ref{eq:M2}) contains a ``hidden" term that depends on the time derivative of the $\vec{B}$ field (usually taken to be zero), but can obscure the electric and magnetic contributions. With further manipulation one can present the modified Maxwell's equations in a similar form to the unmodified equations, given by
\begin{align}
&\vec{\nabla}\cdot\vec{D}_T =\rho_f,\label{eq:M7}\\
&\vec{\nabla} \times \vec{H}_T-\frac{\partial \vec{D}_T}{\partial t}=\vec{J}_f,\label{eq:M8}\\
&\vec{\nabla} \cdot \vec{B}= 0,\label{eq:M9}\\
&\vec{\nabla} \times \vec{E}+\frac{\partial \vec{B}}{\partial t} = 0\label{eq:M10}
\end{align}
with the constitutive relations redefined as
\begin{align}
&\vec{D}_T=\epsilon_0\vec{E}+\vec{P}+\vec{P}_{aB}\text{ where }\vec{P}_{aB}=-g_{a\gamma\gamma}\sqrt{\frac{\epsilon_0}{\mu_0}}(a\vec{B}),\label{eq:M11}\\
&\vec{H}_T=\frac{\vec{B}}{\mu_0}-\vec{M}-\vec{M}_{aE}\text{ where }\vec{M}_{aE}=-g_{a\gamma\gamma}\sqrt{\frac{\epsilon_0}{\mu_0}}(a\vec{E}).\label{eq:M12}
\end{align}
Here $\vec{D}_T$ and $\vec{H}_T$ are the modified definitions of the fields that satisfy equations (\ref{eq:M7}) to (\ref{eq:M10}) due to the additional axion terms, $\vec{P}_{aB}$ and $\vec{M}_{aE}$. 

This description is also similar to that which is adopted when deriving modified Maxwell's equations for Lorentz invariance violations in the photon sector of the Standard Model Extension (SME)~\cite{KM}. The SME includes in the Lagrangian all possible Lorentz invariance violations, and by comparison to the SME it is apparent that $g_{a\gamma\gamma}a$ is similar to an odd-parity Lorentz invariance violation, characterised by the coefficients, $\kappa_{DB}$ or $\kappa_{HE}$. The difference is that in the SME these coefficients are constants, and describe a background DC field related to an absolute frame of reference. In a typical experiment, time variation is induced by rotating with respect to this frame of reference (Michelson-Morley experiments), the equations then become similar to modified axion electrodynamics with the rotation frequency equivalent to the axion Compton frequency. A quasi-static theory has been developed to test odd parity Lorentz-violating using electrostatics and magnetostatics and it has been determined that these effects cannot be suppressed by shielding \cite{Bailey2004}. This type of odd parity Lorentz invariance violation experiment is also discussed in detail in \cite{MT} (presented in SI units), which allowed first limits on some types of Lorentz invariance violations. Furthermore, recently, an experiment that is similar to an odd-parity test of Lorentz invariance violations has been configured as a search for low-mass axions\cite{YMichimura}, and it has also been pointed out that the asymmetry in the response to electric and magnetic fields ($\vec{E}$ and  $\vec{B}$) suggests that the cosmic axion field effectively breaks Lorentz symmetry\cite{CHill2015,CHill2016}.

Later in this paper we also show that the extra terms, $\vec{P}_{aB}$ and $\vec{M}_{aE}$ are also analogous to impressed source fields. In electrodynamics impressed fields are typically added to Maxwell's equations to account for energy conversion from another type of energy (i.e. chemical energy in a battery) into electromagnetic energy\cite{RHbook2012}, and are also known to contribute to the Lorentz force. Thus, the oscillating background fields are also analogous to numerous tiny oscillating permanent electric dipoles and permanent magnetic dipoles induced by the axion, causing an oscillating electromotive force (EMF) and magnetomotive force (MMF) respectively. Indeed, it has been pointed out in the past that when a DC $\vec{B}$-field is applied, the axion modifications can be thought of as a polarization field~\cite{HONG1991197,Jooyoo1990}. However, the case we present here is more general and includes both the applied $\vec{E}$ and $\vec{B}$ fields, which are not necessarily DC.

Recently, others have reinterpreted axion electrodynamics without the initial approximation given by Eqn, (\ref{eq:M2b}) \cite{Ouellet2018,Beutter18}. However, this work still ignores any non-irrotational electric behaviour, by a priori assuming the electric vector potential, and hence the Curl of Eqn.(\ref{eq:M11}), is zero, which then leads to the conclusion that $\vec{J}_{a}$ is non-interacting. Also, when creating a magnetic field, it is not valid to set the impressed electrical current to zero (i.e. set $\vec{J}_{f}=0$), which creates the field when undertaking axion induced field calculations, as is done in reference \cite{Younggeun18}. This is because in this case, $\vec{J}_{f}$, is a source term, not a loss term (and the source of the non-irrotational electric effects ignored in \cite{Ouellet2018,Beutter18}). We show, that when this impressed current is not ignored, there is an extra term due to an axion induced voltage source oscillating at the Compton frequency of the axion, which is also represented by an effective magnetic current and an impressed non-conservative electric field. The magnetic current is needed to properly define the boundary conditions at the surface of the electromagnet, while the impressed non-conservative electric field contributes to the Lorentz force. This is not new physics, all voltage sources add an extra term to the Lorentz force, which is only defined within the boundaries of the voltage source \cite{GriffBook}, and can also be represented as an effective magnetic current at the parallel boundary \cite{RHbook2012}. Moreover, it is the time derivative of this oscillating impressed non-conservative electric field, which sources the axion induced current, and can be easily shown to be equivalent to a polarization or displacement current.

It is straightforward to show that the effective axion current is a polarization or displacement current, and that the continuity equation is satisfied. From equations (\ref{eq:M11}) and (\ref{eq:M6}) one can now write the axion current more generally than eqn.(\ref{eq:M2b}), as
\begin{equation}
\vec{J}_{aB} = \frac{\partial \vec{P}_{aB}}{\partial t} = -g_{a\gamma\gamma}\sqrt{\frac{\epsilon_0}{\mu_0}}\frac{\partial (a\vec{B})}{\partial t}. \label{eq:C2} 
\end{equation}
Then from equations (\ref{eq:M5}) we may define the axion induced volume charge density as,
\begin{equation}
\rho_{aB} = g_{a\gamma\gamma}\sqrt{\frac{\epsilon_0}{\mu_0}} \vec{\nabla}\cdot (a\vec{B}), \label{eq:C1} 
\end{equation}
In the situation where there are no free charges or free conducting electrons, we interpret $\rho_{aB}$ as a background bound charge and $\vec{J}_{aB}$ as a background polarization or displacement current. Taking the time derivative of Eqn.(\ref{eq:C1}) and the divergence of Eqn.(\ref{eq:C2}) we obtain
\begin{equation}
\vec{\nabla}\cdot\vec{J}_{aB} = - \frac{\partial \rho_{aB}}{\partial t}, \label{eq:C6}
\end{equation}
demonstrating that the continuity equation is satisfied. This means we may interpret the effective current density, $\vec{J}_{aB}$, due to the displacement of effective bound charge, $\rho_{aB}$, as an oscillating background polarization, $\vec{P}_{aB}$.

Note, from equation (\ref{eq:M6}) there is also an axion induced bound current associated with the induced magnetization given by
\begin{equation}
\vec{J}_{aE} =\vec{\nabla}\times\vec{M_{aE}}= -g_{a\gamma\gamma}\sqrt{\frac{\epsilon_0}{\mu_0}}\vec{\nabla}\times (a\vec{E}).
\label{eq:C8}
\end{equation}

\subsection{Axion Modifications as Impressed Sources}

In the previous subsection we reformulate the modified equations to highlight the similarity to an odd-parity Lorentz invariance violations, in this section we reformulate the equations to highlight the modifications as impressed sources. The prior analysis has been general, however in this subsection we assume linear magnetic and dielectric systems, such that $\vec{P}=\chi_e\epsilon_0\vec{E}$ and $\vec{M}=\chi_m\epsilon_0\vec{H}$ where $\epsilon_r=1+\chi_e$ and $\mu_r=1+\chi_m$. Then we define \textit{total} electric and magnetic fields such that the following constitutive relationships hold 
\begin{equation}
\vec{E}_T=\frac{1}{\epsilon_r\epsilon_0}\vec{D}_T,\label{Cons1}
\end{equation}
\begin{equation}
\vec{B}_T=\mu_r\mu_0\vec{H}_T,\label{Cons2}
\end{equation}
which lead to the following definitions of total fields,
\begin{equation}
\vec{E}_T=\vec{E}+\vec{E}_{aB}, \ \text{where} \ \vec{E}_{aB}=-g_{a\gamma\gamma}\frac{c}{\epsilon_r}(a\vec{B}),\label{ET1}
\end{equation}
\begin{equation}
\vec{B}_T=\vec{B}+\vec{B}_{aE}, \ \text{where} \ \vec{B}_{aE}=g_{a\gamma\gamma}\frac{\mu_r}{c}(a\vec{E}).\label{BT1}
\end{equation}
It is interesting to note that in these redefinitions, $\vec{E}_T$ is comprised of a curl-free component, $\vec{E}$, and the axion induced component, $\vec{E}_{aB}$, which is solenoidal with a non-zero curl, and can be identified as a non-conservative electric field or voltage source, which supplies an EMF. Thus, $\vec{E}_{aB}$ contributes to the Lorentz force, just like any impressed voltage source which supplies an extra non-conservative force per unit charge\cite{TobarElectret}. Likewise $\vec{B}_T$ is comprised of a divergence-free component, $\vec{B}$, and the axion induced component, $\vec{B}_{aE}$, which is diverging and sources an MMF.

To take the next step to obtain the set of modified equations in terms of the total fields, we rearrange the equations (\ref{eq:M7})-(\ref{eq:M10}) such that only the $\vec{E_T}$ and $\vec{B_T}$ fields exist on the left hand side of the modified equations, with the source terms on the right hand side. Now, the first two modified axion electrodynamic equations in terms of the total fields can be found by substituting Eqn.(\ref{Cons1}) into Eqn.(\ref{eq:M7}) to give Eqn.(\ref{Im1}) and substituting Eqn.(\ref{Cons2}) into Eqn.(\ref{eq:M8}) to give Eqn.(\ref{Im2}). 
\begin{align}
&\vec{\nabla}\cdot\vec{E}_T =\frac{\rho_f}{\epsilon_r\epsilon_0},\label{Im1}\\
&\vec{\nabla} \times \vec{B}_T-\frac{\epsilon_r\mu_r}{c^2}\frac{\partial \vec{E}_T}{\partial t}=\mu_r\mu_0\vec{J_f},\label{Im2}
\end{align}
To derive the last two modified axion equations of the four (i.e. Eqn.(\ref{Im3}) and Eqn.(\ref{Im4})), we take the divergence of Eqn.(\ref{BT1}) and the curl of Eqn.(\ref{ET1}), then implement well known vector identities and assume $\vec{\nabla}a$ is zero. Then it may be shown that to first order in $g_{a\gamma\gamma}a$ the modified magnetic Gauss' law and the modified Faraday's law become (see appendix),
\begin{align}
&\vec{\nabla} \cdot \vec{B}_T=g_{a\gamma\gamma}a\frac{c}{\epsilon_r}\mu_r\mu_0\rho_f,\label{Im3}\\
&\vec{\nabla} \times \vec{E}_T+\frac{\partial \vec{B}_T}{\partial t} =-g_{a\gamma\gamma}a\frac{c}{\epsilon_r}\mu_r\mu_0\vec{J_f}.\label{Im4}
\end{align}

In a source free medium there are no currents or charges adding energy to the system. For this case, the above equations only describe a dissipative system where electromagnetic energy is lost, or converted to heat (one could also model this as complex $\vec{E}$ and $\vec{B}$ fields due to losses in materials). To describe the excitation of the system with energy from another source, we require the inclusion of impressed sources (as discussed in Harrington\cite{RHbook2012}). If a system is lossless and has zero net charge we can in general assume $\rho_f=0$, and any conduction currents are zero. To excite the system with impressed source terms is more favourable to impress a magnetic field via an electrical current such that $\vec{J}_f =\vec{J}_f^i$. We assume this situation in the following sections.

\section{Axion Induced Oscillating Sources and Fields under a DC Magnetic field}

A commonly proposed method to increase experimental sensitivity to axions is to drive a strong electric current through the coil of a  DC electromagnet to drive a large DC magnetic field. Also, magnetic materials can be unacceptably lossy, so we confine the following analysis to lossless dielectrics. Thus, we assume $\mu_r=1$ (i.e. non-magnetic materials), $\vec{J}_f^c=0$ and $\rho_f=0$ (i.e. no electric conduction or volume losses). To create a large DC magnetic field we apply a large DC current source to the coil of an electromagnet such that $\vec{J}_f^i=\vec{J}_{f_0}^i$, which is described as an impressed current source (created through an external source of energy, e.g. mains power). Also, due to no directly applied irrotational electric field, then $\vec{E}\approx0$ (of order $g_{a\gamma\gamma}^2a^2$) so that $\vec{B}_{aE}$ is also essentially zero and can be ignored. Under these assumptions the modified axion equations (\ref{Im1}-\ref{Im4}) become,
\begin{align}
&\vec{\nabla}\cdot\vec{E}_T=0,\label{Im5}\\
&\vec{\nabla} \times \vec{B}-\frac{\epsilon_r}{c^2}\frac{\partial \vec{E_T}}{\partial t}=\mu_0\vec{J}_{f_0}^i,\label{Im6}\\
&\vec{\nabla} \cdot \vec{B}=0,\label{Im7}\\
&\vec{\nabla} \times \vec{E}_T+\frac{\partial \vec{B}}{\partial t} =-g_{a\gamma\gamma}a\frac{c}{\epsilon_r}\mu_0\vec{J}_{f_0}^i=-\vec{J}_{ma}^i.\label{Im8}
\end{align}
Thus, by using the Weber convention, we see that the impressed DC current density, $\vec{J}_{f_0}^i$, induces an impressed magnetic current (or voltage source) oscillating at the axion Compton frequency through the inverse Primakoff effect. This is given by
\begin{equation}
\vec{J}_{ma}^i=g_{a\gamma\gamma}a\frac{c}{\epsilon_r}\mu_0\vec{J}_{f_0}^i.
\label{MC}
\end{equation}
Note that the term $\frac{\partial \vec{B}}{\partial t}$ in the modified Faraday's equation can be defined as the magnetic displacement current\cite{RHbook2012}. 

The effective axion magnetic current is due to the input of external energy into the system, which causes the background bound charge to oscillate due to the impressed energy source created from the inverse Primakoff effect. This fact means there is a non-zero electric curl, which creates an electromotive force in a similar way to any voltage source, as outlined in standard electrodynamics text books, such as Griffiths \cite{GriffBook}. This non-conservative force per unit charge, $\vec{f}_{aB}=\vec{E}_{aB}$, may be considered as an impressed electric field \cite{RHbook2012,Balanis2012}, or a permanent impressed polarization, $\vec{P}_{aB}$ \cite{TobarElectret}, and from the curl of Eqn. (\ref{ET1}) is related to the magnetic current by. 
\begin{equation}
\vec{\nabla} \times \vec{f}_{aB}=\vec{\nabla} \times \vec{P}_{aB}/(\epsilon_r\epsilon_0)=-\vec{J}_{ma}^i=-g_{a\gamma\gamma}a\frac{c}{\epsilon_r}\mu_0\vec{J}_{f_0}^i
\label{emf}
\end{equation}
Then via, Stoke's theorem, in integral form, the electromotive force may be calculated from,
\begin{equation}
\mathcal{E}=\oint_P\vec{f}_{aB}\cdot d\vec{l}=-g_{a\gamma\gamma}a\frac{c}{\epsilon_r}\mu_0\int_S\vec{J}_{f_0}^i\cdot d\vec{a},
\end{equation}
Defining $d$ as the length around the enclosed path of the integration, and given that $I_{f_0,enc}^i=\int_S\vec{J}_{f_0}^i\cdot d\vec{a}$, gives
\begin{equation}
\mathcal{E}=E_{aB}d=-I_{ma \ enc}^i=-g_{a\gamma\gamma}a\frac{c}{\epsilon_r}\mu_0I_{f_0 \ enc}^i.
\end{equation}
Thus, the impressed electric field, $E_{aB}=\mathcal{E}/d$, may be considered as an EMF per unit length, and is related to the enclosed magnetic current, $I_{ma \ enc}^i$, via a left hand rule (hence minus sign).

For such voltage sources, the effective magnetic current also defines the parallel boundary condition of the impressed electric field \cite{TobarElectret} (shown more rigorously later in this paper). However, for this case, the nature is different to a typical voltage source because it resembles a magnetoelectric effect where the oscillating charge does not reside on the exterior perpendicular surface like in a standard voltage source, which terminates the impressed electric field within a finite volume. The dynamics is such that oscillating axion bound charge exists even though the effective net charge in the volume is zero and there is no perpendicular surface. This situation has equal parts of the background charge with opposite sign oscillating simultaneously in opposite directions within the volume, contributing to a non-zero oscillating displacement current or polarization field but with a zero net charge. Such dynamics is best described by the effective impressed magnetic current source, resulting in the axion induced impressed electric field (or force per unit charge), which can be defined through an electric vector potential rather that a scalar electric potential. The dynamics of similar systems in electrodynamics has also been solved through the two-potential formulation of electrodynamics, which is introduced in the next section.  

We also point out that an experiment, which is sensitive to this electric vector potential will be directly sensitive to the static value of the fundamental parameter, $\theta(t)=a/f_a$. It has been pointed out in reference\cite{Cao2017} that this is indeed possible, depending on the nature of the boundary source. In this work we identify the main boundary source as the axion induced impressed magnetic current.

\section{Two-Potential Formalism}

It is evident that Maxwellian electrodynamics with impressed sources is of similar form to axion modified electrodynamics. For example, both describe an extra non irrotational (or non conservative) electric field impressed into the equations. Thus, in a similar way, the modified electrodynamic equations in terms of the total fields given by Eqns (\ref{Im1})-(\ref{Im4}) require a two-potential solution similar to those given by \cite{RHbook2012,Balanis2012,Cabibbo1962,Keller2018}. The two potential formulation is used in electrodynamics and antenna theory to model voltage sources (non conservative electric fields), when there is conversion of external energy into electromagnetic (such as from mains power, or the chemistry of a battery) \cite{RHbook2012,Balanis2012}. The fact is that in many cases, a distribution of electric charge can be mathematically replaced by an equivalent distribution of magnetic current when there is a conversion of external energy into electromagnetic energy.  This concept has also been useful in numerical electromagnetic analysis, when presented as a two-potential formalisation \cite{Kudryavtsev2013}. Also, recently the effective magnetic current technique was applied to oscillating bound charge sources within Maxwellian electrodynamics, such as an electret with an oscillating permanent polarization\cite{TobarElectret}. We must stress that there are no magnetic monopoles represented in these equations. 

Applying this technique to Eqns. (\ref{Im5}-\ref{Im8}), the divergences of the total fields are zero, so we simplify the analysis by applying the two vector potential formulation. In general, the source terms are magnetic and electric current density, with the former in the same units of magnetic flux density per unit time, induced via equation (\ref{MC}) as discussed in the previous subsection. Thus, by using the two vector potential formulation and the principle of superposition, we may consider a general impressed electric, $\vec{J}_e^i$, and magnetic, $\vec{J}_m^i$, current sources separately via the following equations.
\begin{equation}
\vec{\nabla} \times \vec{E}_{A}=-\frac{d\vec{B}_{A}}{dt} \ \text{and} \ \vec{\nabla} \times \vec{B}_{A}=\frac{\epsilon_r}{c^2}\frac{d\vec{E}_{A}}{dt}+\mu_0\vec{J}_e^i
\label{elecJ}
\end{equation}
\begin{equation}
\vec{\nabla} \times \vec{E}_{C}=-\frac{d\vec{B}_{C}}{dt}-\vec{J}_m^i  \ \text{and} \  \vec{\nabla} \times \vec{B}_{C}=\frac{\epsilon_r}{c^2}\frac{d\vec{E}_{C}}{dt}
\label{magJ}
\end{equation}
and using the superposition principle the total fields may be written as,
\begin{equation}
\vec{E}_T=\vec{E}_{A}+\vec{E}_{C}
\end{equation}
\begin{equation}
\vec{B}=\vec{B}_{A}+\vec{B}_{C}
\end{equation}
Solving the electric current case, we use the usual magnetic vector potential $\vec{A}$, where $\vec{B}_{A}=\vec{\nabla} \times \vec{A}$, which will allow the calculation of $\vec{E}_A$ and $\vec{B}_A$. Solving the magnetic current case, we use the electric vector potential, $\vec{E}_{C}=-\frac{1}{\epsilon_0\epsilon_r}\vec{\nabla} \times \vec{C}$, which will allow the calculation of $\vec{E}_C$ and $\vec{B}_C$.

Thus, the non-irrotational electric behaviour contributes a vector potential term, which is not considered in \cite{Ouellet2018,Beutter18} (effectively these reference ignore the non-zero curl of $\vec{E}_{aB}$), so that \cite{RHbook2012},
\begin{equation}
\vec{C}( \vec{r},t )=\frac {\epsilon _ {0}\epsilon _ {r} } {4\pi} \int _ {\Omega } \frac{ \vec{J}_{ma}^i \left( \vec{ r } ^ { \prime } , t ^ { \prime } \right) } { \left| \vec{ r }-\vec{ r } ^ {\prime} \right| } \mathrm{d}^{3} \vec{r}^{\prime}.
\end{equation}
Then by substituting in eqn(\ref{MC}), we obtain
\begin{equation}
\vec{C}( \vec{r},t )=\frac {g_{a\gamma\gamma} a }{ 4\pi c} \int_{\Omega } \frac {\vec{J}_{f_0}^i \left( \vec{ r } ^ { \prime } , t ^ { \prime } \right) } { \left| \vec{r}-\vec{r} ^ { \prime } \right| } \mathrm{d}^{3}\vec{r }^{\prime}.
\end{equation}
The electric field is then related to the electric vector potential by, 
\begin{equation}
\vec{E}_{C}=\vec{E}_{aB}=-\frac{1}{\epsilon_0\epsilon_r}\vec{\nabla}\times\vec{C},
\end{equation}
so that,
\begin{equation}
\vec{E}_{aB}=-\frac{g_{a\gamma\gamma} a c}{\epsilon_r}\frac {\mu_0}{ 4\pi} \int_{\Omega } \frac {\vec{\nabla} \times \vec{J}_{f_0}^i \left( \vec{ r } ^ { \prime } , t ^ { \prime } \right) } { \left| \vec{r}-\vec{r} ^ { \prime } \right| } \mathrm{d}^{3}\vec{r }^{\prime}
\label{EVP}
\end{equation}
Here $\vec{C}$ at point $\vec{r}$ and time $t$ is calculated from magnetic currents at distant position $\vec{r}^{\prime } $ at an earlier time $t^{ \prime }=t-\left|\vec{r}-\vec{r}^{\prime}\right|/c$ (known as the retarded time). The location $\vec{r}^{ \prime } $ is a source point within volume $\Omega$ that contains the magnetic current distribution. The integration variable, ${d} ^{3}\vec{r}^ { \prime }$, is a volume element around position $r^{\prime}$. Note that equation (\ref{EVP}) is the same as equation (37) in \cite{Ouellet2018}, but not just valid for $L>>\lambda$ as suggested in \cite{Ouellet2018}, this calculation shows it is in fact valid over all length scales. The derivation in \cite{Ouellet2018} ignores that $\vec{E}_{aB}$ has a non-zero Curl, and a priori assumes it is zero in the derivation by splitting the derivation of the $\vec{E}_T$-field into the solenoidal ($\vec{E}_{aB}$ term) and irrotational components, and thus leaves it out in the calculation.

For an ideal electromagnet with an effective electric surface current distribution of $\vec{\kappa}_{f0}^i$, the above integral turns into a surface integral of the form,
\begin{equation}
\vec{C}( \vec{r},t )=\frac {g_{a\gamma\gamma} a} {4\pi c} \int _ {S_1} \frac{ \vec{\kappa}_{f0}^i \left( \vec{ r } ^ { \prime } , t ^ { \prime } \right) } { \left| \vec{ r }-\vec{ r } ^ {\prime} \right| } \mathrm{d}^{2} \vec{r}^{\prime},
\end{equation}
\begin{equation}
\vec{E}_{aB}=-\frac{g_{a\gamma\gamma} a c}{\epsilon_r}\frac {\mu_0}{ 4\pi} \int _ {S_1} \frac{ \vec{\nabla}\times\vec{\kappa}_{f0}^i \left( \vec{ r } ^ { \prime } , t ^ { \prime } \right) } { \left| \vec{ r }-\vec{ r } ^ {\prime} \right| } \mathrm{d}^{2} \vec{r}^{\prime},
\label{EVP2}
\end{equation}
where, $S_1$ defines the surface over which the electric current flows.

\section{Quasi-static Approximation for Low-Mass Axion Detection under DC Magnetic Field}

In electrodynamics quasi-static approximations are justified if time rates of change are slow enough (i.e. frequencies are low enough) so that time delays due to the propagation of photons are unimportant. This occurs when the wavelengths of the photons are much larger than the size of the electrodynamic system under consideration (i.e. sub-wavelength). This is certainly the case for circuits designed from lumped elements for applications at low frequencies, and electromagnetic effects from low-mass axions in typical experimental contexts. In this regime there are no propagating fields and solutions are basically in the near field regime. Typically, these systems are analysed in terms of voltage and current sources combined with circuit components such as resistors, capacitors, and inductors. For axion modified electrodynamics, if the axion mass is below 1 $\mu eV$ a photon frequency of less than 240 MHz will be produced under a DC magnetic field. This is equivalent to a wavelength of more than $1.2~m$, whereas the bore of a high field magnet is typically on the order of $10~cm$. Thus, searches for axions at these mass scales can be considered to be in the quasi-static limit. In this limit the solutions are not in the form of a propagating wave, but only in terms of time dependent oscillating fields.

When applying a DC magnetic field, for low-mass axions as defined above, we can implement the standard Magneto quasi-static (MQS) technique. To calculate the first terms using this approximation, Ampere's law is modified and the displacement current is set to zero such that $\frac{d\vec{E}}{dt}=0$, and hence $\vec{\nabla}\times\vec{B}\approx \mu J_f$. For modified axion electrodynamics, this means we should assume $\frac{d\vec{E}_T}{dt}=0$ to calculate the first approximation. For this case we have $\vec{B}=\vec{B}_0$, $\frac{d\vec{B}}{dt}=0$, $\vec{E}=0$ and $\vec{E}_T=\vec{E}_{aB}=-g_{a\gamma\gamma}\frac{c}{\epsilon_r}(a\vec{B_0})$ \cite{McAllisterFormFactor}. Assuming these values and $\vec{\nabla}a=0$, equations (\ref{Im5}) to (\ref{Im8}) become,
\begin{align}
&\vec{\nabla}\cdot\vec{E}_T=-g_{a\gamma\gamma}\frac{c}{\epsilon_r}\vec{\nabla}\cdot(a\vec{B_0})=0,\label{QS1}\\
&\vec{\nabla} \times \vec{B}=\vec{\nabla} \times \vec{B}_0=\mu_0\vec{J}_{f_0}^i,\label{QS2}\\
&\vec{\nabla} \cdot \vec{B}=\vec{\nabla} \cdot \vec{B}_0=0,\label{QS3}\\
&\vec{\nabla} \times \vec{E}_T=-g_{a\gamma\gamma}\frac{c}{\epsilon_r}\vec{\nabla} \times(a\vec{B_0})=-g_{a\gamma\gamma}a\frac{c}{\epsilon_r}\mu_0\vec{J}_{f_0}^i.\label{QS4}
\end{align}
All the above equations are satisfied in the quasi static limit, so the first order solution is,
\begin{equation}
\vec{B}=\vec{B}_0\ \ \text{and} \ \ \vec{E}_T=\vec{E}_{aB}=-g_{a\gamma\gamma}\frac{c}{\epsilon_r}a\vec{B_0}, 
\label{EMF_a}
\end{equation}
which is the same as the particular solution derived by Hill \cite{CHill2015,CHill2016}. More generally, this solution is correct in the DC or low-mass limit if the time rate of change is small. As the time rate of change increases, more terms of the series are required to accurately calculate the $\vec{B}$ and $\vec{E}_T$ fields. To calculate the next term we assume $\vec{B}=\vec{B}_0+\vec{B}_a(\omega_a)$, where $\vec{B}_a$ will be the first term of the axion induced time varying magnetic field at frequency $\omega_a$. Given that $\vec{E}=0$ and $\vec{E}_T=\vec{E}_{aB}=-g_{a\gamma\gamma}\frac{c}{\epsilon_r}a\vec{B}_0$, we obtain the following relationship for the axion induced magnetic field by considering only the time varying components from equation (\ref{Im6}).
\begin{equation}
\vec{\nabla} \times\vec{B}_a=\frac{\epsilon_r}{c^2}\frac{d\vec{E}_{aB}}{dt}=-\frac{g_{a\gamma\gamma}}{c}\frac{da}{dt}\vec{B}_0=\mu_0\vec{J}_{a}.
\label{amp}
\end{equation}

This is the standard equation used to calculate the oscillating axion induced magnetic fields for low mass experiments~\cite{Sikivie2014a,ABRACADABRA}. Note the value of $\vec{B}_a$ above, which is calculated from the time derivative of $\vec{E}_{aB}$, is suppressed by a factor of the Compton frequency with respect to $\vec{E}_{aB}$ in the low-mass (and hence low-frequency) regime. Thus, we can conclude that the effective axion current, $\vec{J}_a$ (see Eq. (\ref{eq:M2b}) and (\ref{eq:C2})) is a displacement current sourced by the time derivative of the time varying field, $\vec{E}_{aB}$. Following this, the next correction to $\vec{E}_T$ can be calculated through the time derivative of $\vec{B}_a$. This term will be suppressed by the square of the Compton frequency, so in the low-mass regime it may be ignored.

\subsection{Boundary Conditions}

When solving the modified Maxwell's equations in differential form (Eqns. (\ref{Im5})-(\ref{Im8})), in general, the boundary conditions must also be known. The integral forms may be used to determine the modifications to the standard electromagnetic boundary conditions, and are given below (here $d\vec{a}$ is the infinitesimal vector element of area). 
\begin{equation}
\oiint_S\vec{E}_T\cdot d\vec{a} = 0\label{eq:I1},
\end{equation}
\begin{equation}
\oint_P \vec{B}\cdot d\vec{l}=\mu_0I_{f_0enc}^i+\mu_0\epsilon_r\epsilon_0\frac{d}{dt}\int_S\vec{E}_T\cdot d\vec{a}\label{eq:I2}
\end{equation}
\begin{equation}
\oiint_S \vec{B}\cdot d\vec{a} = 0\label{eq:I3},
\end{equation}
\begin{equation}
\oint_P\vec{E_T}\cdot d\vec{l}=-\frac{d}{dt}\int_S\vec{B} \cdot d\vec{a}-g_{a\gamma\gamma}a\frac{c}{\epsilon_r}\mu_0I_{f_0enc}^i\label{eq:I4}
\end{equation}
Here, $I_{f_0enc}^i= \int_S\vec{J}_{f_0}^i\cdot d\vec{a}$.

From these integral equations it is straight forward to derive the modified boundary conditions between two regions labeled $1$ and $2$ as follows; 
\begin{equation}
\vec{E}_{T1}^{\perp}=\vec{E}_{T2}^{\perp},\label{eq:BC1}
\end{equation}
\begin{equation}
\vec{B}_{1}^{\parallel}-\vec{B}_{2}^{\parallel}=\mu_0\vec{\kappa}_{f_0}^i\times\hat{n},\label{eq:BC4}
\end{equation}
\begin{equation}
\vec{B}_{1}^{\perp}=\vec{B}_{2}^{\perp}.\label{eq:BC2}
\end{equation}
\begin{equation}
\vec{E}_{T1}^{\parallel}-\vec{E}_{T2}^{\parallel}=-g_{a\gamma\gamma}a\frac{c}{\epsilon_r}\mu_0\vec{\kappa}_{f_0}^i\times\hat{n}=-\vec{\kappa}_{m}^i\times\hat{n}=\vec{E}_{aB},\label{eq:BC3}
\end{equation}
Here $\vec{\kappa}_{f_0}^i$ is the impressed free DC surface current at the boundary and $\vec{\kappa}_{m}^i$ is the impressed surface magnetic current at the boundary oscillating at the axion Compton frequency, and $\hat{n}$ is the normal to the surface on which the surface current flows. 

\subsection{DC Solenoid of Infinite Length}

\begin{figure}[t]
\includegraphics[width=1.0\columnwidth]{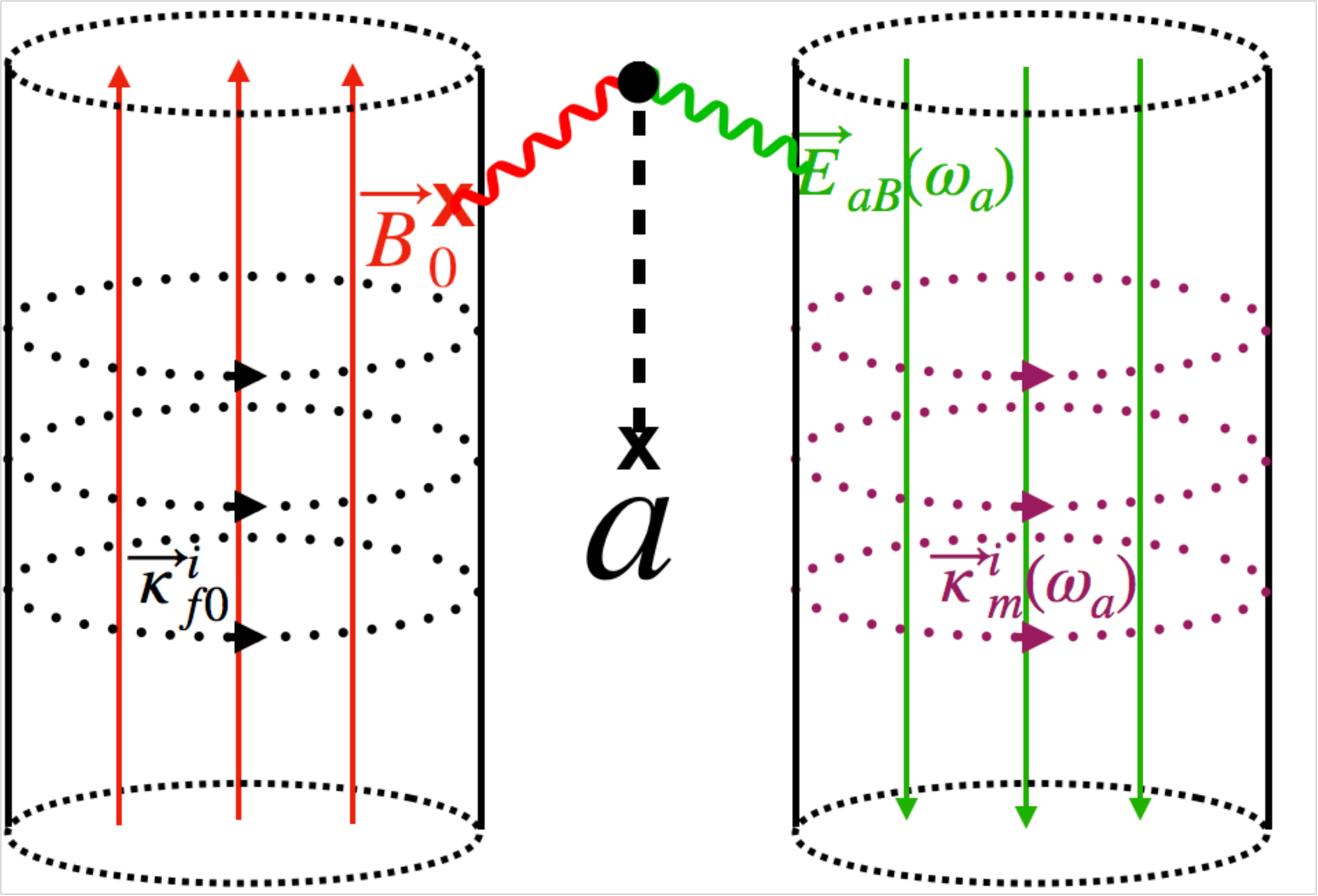}
\caption{Schematic of the axion two-photon interaction inside an ideal infinite solenoid. Left, an ideal DC magnetic field of $\vec{B}_0=B_0\hat{z}$ induced by an ideal impressed surface current of $\vec{\kappa}_{f_0}^i$, nominally produced from an external ideal power supply. Right, resulting impressed magnetic current, $\vec{\kappa}_{m}^i=g_{a\gamma\gamma}a\frac{c}{\epsilon_r}\mu_0\vec{\kappa}_{f_0}^i$ and associated impressed electric field, $\vec{E}_{aB}=-g_{a\gamma\gamma}a_0\frac{c\vec{B}_{0}}{\epsilon_r}e^{-j\omega_a t}\hat{z}$, oscillating at the axion Compton frequency due to the inverse Primakoff effect.}
\label{Inf}
\end{figure}

Currently the most common axion haloscopes apply a strong DC magnetic field to covert the mass of axions to photons of equivalent frequency. In typical experiments, a solenoid electromagnet is aligned in the laboratory $\hat{z}$ direction with an impressed DC free current idealised to be of the form $\kappa_{f_0}^i\hat{\phi}$ at radius $r=R$ (here $R$ is the radius of the solenoid). Thus, the generated fields are ideally equivalent to $\vec{E}=0$ and $\vec{B}=B_0 \hat{z}$ within the cylindrical solenoidal magnet \cite{Wuensch,hagmann1990,Bradley2003,ADMXaxions2010,ADMX2011,Sikivie2014a,7390193,McAllisterFormFactor,McAllister:2016fux,Klash,JEONG2018412}. To first order the system may be approximated as an infinite solenoid as shown in Fig.\ref{Inf}. In the case of the MQS solution discussed in the prior sections, this implies that $\vec{B}=B_0 \hat{z}+\vec{B}_a(\omega_a)$ and $\vec{E}_T=\vec{E}_{aB}(\omega_a)=-g_{a\gamma\gamma}a\frac{cB_0}{\epsilon_r}\hat{z}$. Given that the complex axion scalar field can be written as $a=a_0e^{-j\omega_a t}$ the value of $\vec{E}_{aB}(\omega_a)$ will be,
\begin{equation}
\vec{E}_{aB}(\omega_a)=-g_{a\gamma\gamma}a_0\frac{cB_0}{\epsilon_r}e^{-j\omega_a t}\hat{z}. \label{eq:Ea}
\end{equation}

The normal boundary conditions for $\vec{B}$ and $\vec{E}_T$ must be continuous due to equations (\ref{eq:BC1}) and (\ref{eq:BC2}). However, the parallel boundary equations (\ref{eq:BC3}) and (\ref{eq:BC4}) must be applied, where the subscript ``in" refers to inside the solenoid and the subscript ``out" refers to outside the solenoid (as shown in Fig.\ref{Inf}). Matching both the DC and AC terms (at the axion Compton frequency) separately, from the MQS solutions, and equations (\ref{eq:BC3}) and (\ref{eq:BC4}), we obtain the following:\\
1) DC $\vec{B}$-field boundary condition,
\begin{equation}
\vec{B}_{0~in}^{\parallel}=\mu_0\vec{\kappa}_{f_0}^i\times\hat{n}=\vec{B}_{0}\hat{z}, \ \text{and} \ \vec{B}_{0~out}^{\parallel}=0\label{solBC1}
\end{equation}
2) AC $\vec{E}$-field boundary condition,
\begin{equation}
\vec{E}_{aB}^{\parallel}(\omega_a)_{in}=-g_{a\gamma\gamma}a\frac{c}{\epsilon_r}\mu_{0}\vec{\kappa}_{f_0}^i\times\hat{n}=-g_{a\gamma\gamma}a_0\frac{c\vec{B}_{0}}{\epsilon_r}e^{-j\omega_a t}\hat{z},\label{solBC2}
\end{equation}
\begin{equation}
\\ \vec{E}_{aB}^{\parallel}(\omega_a)_{out}=0\label{solBC3}
\end{equation}
3) AC $\vec{B}$-field boundary condition,
\begin{equation}
\vec{B}_{a}^{\parallel}(\omega_a)_{in}=\vec{B}_{a}^{\parallel}(\omega_a)_{out}.\label{solBC4}
\end{equation}
Thus, inside the solenoid $\vec{E}_{aB}(\omega_a)$ is given by Eqn. (\ref{eq:Ea}), and
\begin{equation}
\frac{d\vec{E}_{aB}}{dt}(\omega_a)=j\omega_ag_{a\gamma\gamma}a_0\frac{c\vec{B}_{0}}{\epsilon_r}e^{-j\omega_a t}\hat{z} 
\end{equation}
There will be no directly induced axion $\vec{B}_a$-field as $\vec{E}=0$. Instead an orthogonal $\vec{B}_a$-field, $B_{a}\hat{\phi}$ must be induced by the oscillating $\vec{E}_{aB}$-field, which may be calculated from the integral form of the modified Ampere's Law  (given by Eqn.(\ref{eq:I2})). Assuming no free current oscillating at frequency $\omega_a$ (there is only DC free current), this may be written as
\begin{equation}
\oint_C B_a(\omega_a)\hat{\phi}\cdot d\vec{l} = \frac{\epsilon_r}{c^2}\int\int_S \frac{d\vec{E}_{aB}}{d t}(\omega_a)\cdot d\vec{a}.\label{eq:BaI}
\end{equation}

In many cases, the infinite solenoid approximation suffices, as the experiment is embedded in the middle of the constant field region of the magnet. Outside the solenoid ($r>R$) the axion induced fields must be calculated from the boundary condition between media given by Eqn.~(\ref{solBC4}) at $r=R$. In this region, outside the solenoid there, are no axion induced magnetizations or polarizations. In modified electrodynamics this must be treated as a different medium. To match the boundary condition between the media, a $\vec{B}$-field must be produced  at the boundary, which can be calculated to be
\begin{equation}
\vec{B}_{\phi a}(\omega_a)  = j\omega_ag_{a\gamma\gamma}a_0\frac{B_0}{c}\frac{R}{2}e^{-j\omega_a t}\hat{\phi}~~(r=R).\label{eq:Inf1}
\end{equation}
Outside the solenoid, the solution is very similar to an infinite current carrying conductor (with a total current of $\vec{I}_a = \vec{J}_a\times\pi R^2$), with an oscillating magnetic field produced. We can calculate the field using the integral form of Eqn.~(\ref{eq:M8}):
\begin{equation}
\vec{B}_{\phi a}(r,\omega_a)  =  j\omega_ag_{a\gamma\gamma}a_0\frac{B_0}{c}\frac{R^2}{2r}e^{-j\omega_a t}\hat{\phi}~~(r>R),\label{eq:Inf2}
\end{equation}
Thus, outside the infinite solenoid it is more sensitive to detect $\vec{B}_a$, but inside the solenoid it is more sensitive to detect $\vec{E}_{aB}$ as proposed in \cite{BEAST}.

\section{Discussion on Background Polarizations and Magnetizations}

Since the axion modifications appear as impressed source terms in Maxwell's equations, it is instructive to think of the axion effects as oscillating background bound charges and currents. As an aside, background vacuum polarization and magnetization effects  cause the running of the fine structure constant, $\alpha$, due to equal components of electric screening (polarization of vacuum) and magnetic anti-screening (magnetization of vacuum). These effects cause the perceived quantum of electric charge to increase at small distances, while the perceived quantum of magnetic flux decreases~\cite{WilczekAF,TobarFSC2005}, thus the fine structure constant increases at small distances and high energy scales, with the vacuum effectively acting as a dielectric ($\epsilon_r>1$) and paramagnetic ($\mu_r<1$) medium (which has been confirmed experimentally \cite{FSCrun2000,FSCrun2017}). Furthermore, it has been shown that background effects become dominant in the regime where the length scale of the experiment is much smaller than the Compton wavelength of the particle, which is true for the low-mass axion regime. This is highlighted for example by the Uehling potential from polarized electron-positron pairs~\cite{Uehling}. In actual fact the background fields in QCD are more complicated than simple electron-positron pairs, which are simply presented here as an example. Thus, the oscillating magnetization and polarization could be interpreted as tiny oscillations of the fine structure constant due to oscillations in the screening and anti-screening processes. Furthermore, as with SME modified electrodynamics, electromagnetic shielding will not suppress axion signals that can be directly detected from the oscillating vacuum polarization and magnetization fields, as they are source terms that generate oscillating EMFs and MMFs respectively. 

One might also expect that a fluctuating fine structure constant could be measurable through variations of the resonant frequency of an appropriately designed resonant cavity or circuit system. This would be analogous to a dielectric resonant cavity effected by Brillouin scattering in the media, induced by refractive index fluctuations, which is also an inherently non-linear process. Recently a dual-mode pumped resonant system has been analysed and shown to be sensitive to axion-induced frequency shifts. Under the appropriate conditions such a system has shown to be sensitive enough to place limits on popular axion models~\cite{freqmetrology}. 

\section{Conclusion}

In this work we have reformulated the modified electrodynamics for the QCD axion, retaining a familiar form to the non-modified Maxwell's equations, with all axion modifications represented in the constitutive relations in a similar way to Lorentz invariance violations. This leads to the identification of oscillating background polarization and magnetization induced by axion conversion under strong DC magnetic and electric fields, which are directly proportional to the axion's scalar amplitude. We show that these fields are analogous to tiny oscillating dipole permanent electrets and magnets respectively, which can be considered as impressed voltage and current sources, representing a conversion of axion mass energy into electromagnetic energy. 

We have also defined the appropriate boundary conditions that should be applied in regions where axion conversion takes place and in regions where it does not. In particular, we show that a DC current, which drives a magnet (and defines the boundary condition of a solenoidal magnetic field) is converted through the inverse Primakoff effect to a parallel, axion induced effective magnetic current oscillating at the axion Compton frequency. This effective magnetic current creates an axion induced oscillating electromotive force, which is determined by an electric vector potential rather than a magnetic vector potential.  However, the central difference is that the axion modifications inside a solenoid cause a polarization, which is more similar to a magneto-electric effect. Thus, for the axion modified electrodynamics there are no normal surface discontinuities to the polarization field, as it is continuous like the applied magnetic field which induced it. However, oscillating bound charges do exist such that the associated polarization field (or force per unit charge) is purely solenoidal and divergence free, similar to the magnetic field driving it and best represented by the effective magnetic current derived at the boundary. This field is a non-conservative electric field, where the integral around the closed path is non-zero, and modifies the Lorentz force in a similar way to how a standard voltage source modifies the Lorentz force\cite{GriffBook}.

The consequence of this work shows that it will be more effective to measure observable electric effects from the axion induced voltage source inside a large DC electromagnet
where DC B-field is present\cite{BEAST}. In contrast, in regions outside the DC B-field it is more effective to measure axion induced magnetic effects. However, the axion induced magnetic effects outside the DC electromagnet are suppressed at low-masses by the Compton frequency of the axion, compared to the axion induced electric effects in regions inside the DC B-field. This is contrary to what is concluded in~\cite{Ouellet2018}, as this work only considers effects from the axion induced electric scalar potential and not the axion induced electric vector potential.

\section{Acknowledgements}

This work was funded by Australian Research Council grant No. DP190100071 and CE170100009, the Australian Government's Research Training Program, and the Bruce and Betty Green Foundation. We acknowledge many positive discussions with Professor Ian McArthur and Dr. Alex Millar. We also acknowledge thought provoking discussions with Jonathan Ouellet, Kent Irwin and Aaron Chou and thank Frank Wilczek for thoughts and discussions.

\appendix

\section{Derivation of Axion Modified Magnetic Gauss' Law and Faraday's Law due to Impressed Sources}

In this appendix we derive equation (\ref{Im3}), the modified Magnetic Gauss' Law and equation (\ref{Im4}), the modified Faraday's Law, due to the inverse Primakoff effect and for the case of a linear dielectric and magnetic material.

\subsection{Modified Magnetic Gauss' Law}

Taking the divergence of equation (\ref{BT1}), we obtain;
\begin{equation}
\vec{\nabla}\cdot\vec{B}_T=\vec{\nabla}\cdot\vec{B}+g_{a\gamma\gamma}\frac{\mu_r}{c}\vec{\nabla}\cdot(a\vec{E}).\label{BT2}
\end{equation}
Given that $\vec{\nabla}\cdot\vec{B}=0$ and $\vec{\nabla}\cdot(a\vec{E})=\vec{E}\cdot\vec{\nabla}a+a(\vec{\nabla}\cdot\vec{E})$, then assuming $\vec{\nabla}a=0$ we obtain;
\begin{equation}
\vec{\nabla}\cdot\vec{B}_T=g_{a\gamma\gamma}\frac{\mu_r}{c}a(\vec{\nabla}\cdot\vec{E}).\label{BT3}
\end{equation}
From equation (\ref{eq:M5}) and assuming a linear dielectric, we obtain;
\begin{equation}
\vec{\nabla}\cdot\vec{E} =\frac{\rho_f}{\epsilon_r\epsilon_0}+ g_{a\gamma\gamma} \frac{c}{\epsilon_r}(\vec{\nabla}\cdot (a\vec{B})).\label{BT4}
\end{equation}
Thus, by substituting equation (\ref{BT4}) into equation (\ref{BT3}) and taking first order in $g_{a\gamma\gamma}a$, we obtain;
\begin{equation}
\vec{\nabla} \cdot \vec{B}_T=g_{a\gamma\gamma}a\frac{c}{\epsilon_r}\mu_r\mu_0\rho_f.
\end{equation}
The same as equation (\ref{Im3}).

\subsection{Modified Faraday's Law}

Taking the curl of equation (\ref{ET1}), we obtain;
\begin{equation}
\vec{\nabla}\times\vec{E}_T=\vec{\nabla}\times\vec{E}-g_{a\gamma\gamma}\frac{c}{\epsilon_r}\vec{\nabla}\times(a\vec{B}).\label{ET2}
\end{equation}
Given that $\vec{\nabla}\times\vec{E}=-\frac{\partial B}{\partial t}$ and $\vec{\nabla}\times (a\vec{B})=\vec{\nabla}a\times\vec{B}+a(\vec{\nabla}\times\vec{B})$, then assuming $\vec{\nabla}a=0$ we obtain;
\begin{equation}
\vec{\nabla}\times\vec{E}_T=-\frac{\partial B}{\partial t}-g_{a\gamma\gamma}\frac{c}{\epsilon_r}a(\vec{\nabla}\times\vec{B}).\label{ET3}
\end{equation}
From equation (\ref{eq:M6}) and assuming a linear magnetic material, we obtain;
\begin{equation}
\vec{\nabla}\times\vec{B} =\mu_0\mu_r\vec{J_f} +\frac{\mu_r\epsilon_r}{c^2}\frac{\partial \vec{E} }{\partial t}-g_{a\gamma\gamma}\frac{\mu_r}{c}\left(\frac{\partial (a\vec{B})}{\partial t}+\vec{\nabla}\times (a\vec{E})\right).\label{ET4}
\end{equation}
Thus, by substituting equation (\ref{ET4}) into equation (\ref{ET3}) and taking first order in $g_{a\gamma\gamma}a$, we obtain;
\begin{equation}
\vec{\nabla} \times \vec{E}_T+\frac{\partial \vec{B}_T}{\partial t} =-g_{a\gamma\gamma}a\frac{c}{\epsilon_r}\mu_r\mu_0\vec{J_f}.
\end{equation}
The same as equation (\ref{Im4}).

\end{document}